\documentclass[aps,pra,twocolumn, showpacs]{revtex4}
\usepackage{color, graphicx,hyperref}
\usepackage{amsmath, amssymb}

\newcommand{\NTO}{\begin{smallmatrix}\circ \\ \circ\end{smallmatrix}}

\definecolor{gruen}{rgb}{0,0.7,0.1}
\definecolor{blau}{rgb}{0,0.1,0.7}
\definecolor{rot}{rgb}{0.8,0,0}

\begin{document}

\title{Correlation measurements with on-off detectors}

\author{J. Sperling}\email{jan.sperling2@uni-rostock.de}\affiliation{Arbeitsgruppe Theoretische Quantenoptik, Institut f\"ur Physik, Universit\"at Rostock, D-18051 Rostock, Germany}
\author{W. Vogel}\affiliation{Arbeitsgruppe Theoretische Quantenoptik, Institut f\"ur Physik, Universit\"at Rostock, D-18051 Rostock, Germany}
\author{G. S. Agarwal}\affiliation{Department of Physics, Oklahoma State University, Stillwater, Oklahoma 74078, USA}

\pacs{42.50.Ar, 03.65.Wj,42.50.-p}
\date{\today}

\begin{abstract}
	We present a general method to detect nonclassical radiation fields with systems of on-off detectors.
	We especially study higher order correlations for the identification of nonclassical radiation.
	This allows us to directly characterize quantum correlations by the statistics measured with systems of on-off detectors.
	Additionally, we generalize our method to multiple detector systems for measurements of correlations between light fields.
	We also consider multi-mode radiation fields and isolate nonclassicality in terms of the space time correlations.
	Finally, we present results for the quantum statistics using on-off detectors operating in nonlinear detection modes.
\end{abstract}

\maketitle

\section{Introduction}\label{Sec:1}
	The production of nonclassical radiation and novel measurement devices have become more and more important to test and apply the quantum phenomena of light.
	Hence, a careful measurement analysis is required to distinguish classical correlations from those arising from quantum physics.
	Especially, single photon detection is of fundamental interest.

	The photoelectric detection of radiation is typically given in the quantum version of a Poisson statistics,
	\begin{align}
		p_n=\left\langle{:}\frac{(\eta\hat n)^n}{n!}\exp\left[-\eta\hat n\right]{:}\right\rangle,
	\end{align}
	where ${:}\ldots{:}$ denotes the normal ordering and $\eta$ the quantum efficiency~\cite{MW95,VW06,A13}.
	An identifier for a nonclassical photoelectric counting statistics is given by the Mandel parameter, $Q_{\rm M}=\langle \left(\Delta n\right)^{2}\rangle/\langle n\rangle-1$, in terms of the mean value and the variance~\cite{M79}.
	This second order moment criterion has been generalized to higher order correlations, cf., e.g.~\cite{AT92,SRV05,SV05,V08,MBWLN10}, which can detect quantum correlations beyond those of the Mandel parameter.

	A common method to perform measurements in the single photon regime is done with detector systems, which are based on avalanche photo diodes in the Geiger mode.
	The signal of such detectors is a bit of information: ''click'' or ''no-click'' if photons are detected or not.
	Examples for such systems are detector arrays or (time-bin) multiplexing setups; see, e.g.~\cite{WDSBY04,ASSBW03,FJPF03,ZABGGBRP05,JDC07,LFCPSREPW08,FLCEPW09,ABA10,ALCS10}.

	It has been shown that these detectors have to be described by a quantum version of a binomial statistics~\cite{SVA12a},
	\begin{align}
		c_k{=}\left\langle\!\!{:}\frac{N!}{k!(N-k)!}\!\left(\!\exp\left[-\frac{\eta\hat n}{N}\right]\!\right)^{\!\!N-k}
		\!\left(\!\hat 1{-}\exp\left[-\frac{\eta\hat n}{N}\right]\!\right)^{\!\!k}{:}\!\!\right\rangle,
	\end{align}
	rather than a Poisson form.
	On this basis one can formulate a binomial parameter~\cite{SVA12b},
	\begin{align}
		Q_{\rm B}=N\frac{\langle \left(\Delta c\right)^{2}\rangle}{\langle c\rangle(N-\langle c\rangle)}-1,
	\end{align}
	where the variance and the mean value of the click counting statistics are used.
	If this parameter becomes negative, then the measured system is nonclassical, which recently has been experimentally demonstrated~\cite{BDJDBW13}.
	
	Alternatively, CCD cameras have been used to identify quantum correlations between pixels of the image.
	Surprisingly, EPR correlations can be directly observed in this system~\cite{DSML12a,DSML12b,BDFL08,BDFL10,LBFD08}.
	Moreover, a reconstruction of the state together with its nonclassical features has been done using CCD image sensors~\cite{HPHP05,PHMH12a,PHMH12b,PHMH13}.
	Another kind of photon number resolving detectors is based on superconducting systems, cf. e.g.~\cite{MBGJLMF09}.

	For other detection scenarios a generalization of second order criteria has been done.
	Higher order field and intensity correlations have been described~\cite{SRV05,SV05,MBWLN10,KKV12,KV12,AT92,A93,HM85}.
	Here, it turns out that principal minors of matrix of moments yield a range of applicable conditions.
	Even for space time dependent correlations one can construct such a general approach~\cite{V08}.

	\begin{figure}[ht]
		\vspace*{0.2cm}
		\includegraphics[width=6cm]{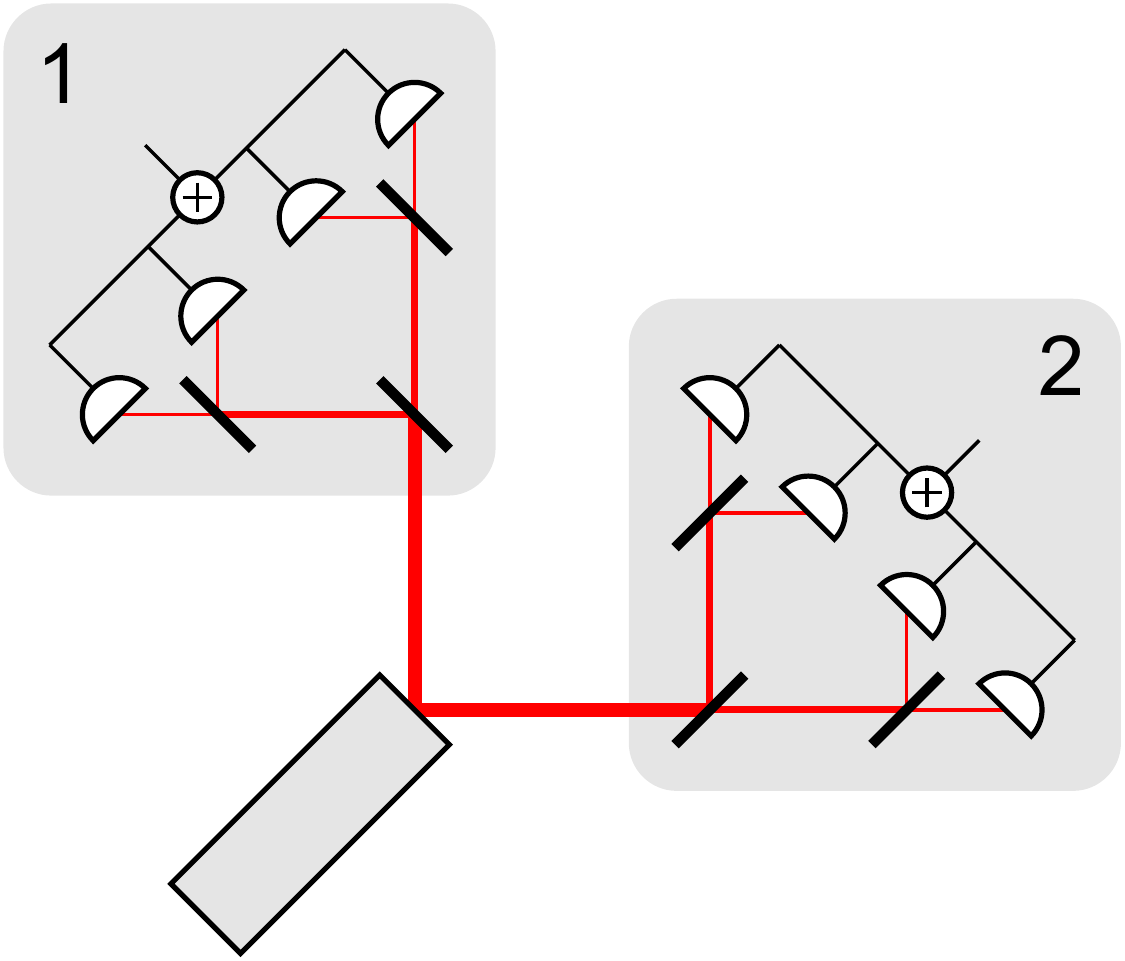}
		\caption{ (Color online)
			The detection of correlated light is schematically shown for $D{=}2$ detectors.
			Light is emitted from a source and split in a multiplexing scheme.
			Both detectors consists of $N_1{=}N_2{=}4$ individual click detectors.
			The individual number of clicks are summed for each detector.
			This results in a joint click counting statistics $c_{k_1,k_2}$.
		}\label{Fig:MultiDet}
	\end{figure}
	In this article, we aim to describe a general measurement scenario; cf. Fig.~\ref{Fig:MultiDet}.
	A source emits nonclassically correlated light in different directions.
	The beams will be detected with one out of $D$ individual on-off detector systems.
	Each of the detector configurations itself consists of $N_d$ ($d=1,\ldots,D$) on-off detectors.
	The coincident click counting statistics $c_{k_1,\ldots,k_D}$ of the joint measurement will be described in such kinds of setups.
	We will derive criteria for nonclassical correlations between the emitted light beams, which are solely based on the measured statistics itself.
	We also study detectors whose response is described through general detection processes.
	For a recent work on the determination of statistics using two photon detectors see~\cite{DYSTS11}.

	We are going to establish a generalized method to identify nonclassical correlations based on higher orders moments of the click-scounting statistics and we will progress to a more general detection process description.
	In Sec.~\ref{Sec:3}, we consider higher order moments of the statistics.
	A matrix of moment criterion will be constructed for the identification of nonclassical radiation.
	We study correlation measurements between multiple detector systems in Sec.~\ref{Sec:3-1}.
	This will allow to detect nonclassical correlations between light fields.
	In Sec.~\ref{Sec:2} we generalize the treatment to arbitrary radiation fields.
	In particular, we deal with multimode radiation fields and isolate nonclassicality in terms of the space time correlations.
	We prove that the binomial form of the click counting statistics will not be changed even for nonlinearly responding detectors.
	We conclude in Sec.~\ref{Sec:4}.

\section{Higher order quantum correlations}\label{Sec:3}
	Let us start with a single mode light field and a linearly responding detector.
	We may define the operator,
	\begin{align}
		\hat \pi=\hat 1-\exp\left[-\tfrac{\eta\hat n}{N}\right].
	\end{align}
	The expectation value of this operator yields the average number $\langle c\rangle$ of clicks divided by the number $N$ of diodes.
	The click counting statistics,
	\begin{align}
		c_k=\left\langle{:}\frac{N!}{k!(N-k)!}\hat \pi^k\left(\hat 1-\hat\pi\right)^{N-k}{:}\right\rangle,
	\end{align}
	for $k\in\{0,\ldots,N\}$, can be fully characterized by its generating function:
	\begin{align}
		g(z)=\sum_{k=0}^N c_k z^k=\langle{:}\left(z\hat\pi+[\hat 1-\hat\pi]\right)^N{:}\rangle.
	\end{align}
	Now, the $m$th factorial moment of the click-counting statistic can be related to the moments $\langle{:}\hat\pi^m{:}\rangle$ (for $m=0,\ldots,N$) by the following relation:
	\begin{align}
		&\sum_{k=0}^N k(k-1)\ldots(k-m+1) c_k\nonumber\\
		=&\left.\frac{\partial^mg(z)}{\partial z^m}\right|_{z=1}
		=\frac{N!}{(N-m)!}\langle{:}\hat\pi^m{:}\rangle.\label{Eq:MomentRelation}
	\end{align}
	Note that the polynomial $x(x-1)\ldots(x-m+1)$, which occurs in this formula, is known as the Pochhammer symbol.
	It uniquely relates the moments of $\hat \pi$ with those of the measured click counting statistics, which are defined as
	\begin{align}
		\langle c^{m}\rangle= \sum_{k=0}^N k^{m} c_k
		\text{ for }m=0,\ldots,N.
	\end{align}
	Moreover, let us stress that Eq.~\eqref{Eq:MomentRelation} shows that the moments $\langle{:}\hat\pi^m{:}\rangle$ can be directly obtained from the measured statistics $c_k$,
	\begin{align}
		\langle{:}\hat\pi^m{:}\rangle=\frac{(N-m)!}{N!}\sum_{k=0}^N k(k-1)\ldots(k-m+1) c_k.
	\end{align}
	This means that no artificial data processing is needed.

	Now, we may consider an operator function $\hat f$, which is based on these moments,
	\begin{align}
		\hat f=\sum_{m=0}^{\lfloor N/2 \rfloor} f_m \hat\pi^m,
	\end{align}
	where the floor function yields $\lfloor N/2 \rfloor=N/2$ for even $N$ and $\lfloor N/2 \rfloor=(N-1)/2$ for odd $N$.
	The floor function is needed to bound the number of moments from above by $N$ in the following steps.
	Immediately, one gets conditions which must be fulfilled for classical states:
	\begin{align}
		0\leq& \langle{:}\hat f^\dagger\hat f{:}\rangle=\sum_{m,m'=0}^{\lfloor N/2\rfloor} f_m^\ast f_{m'}\langle{:} \hat \pi^{m+m'}{:}\rangle.
	\end{align}
	Equivalently, all principal minors of the matrix of moments $\boldsymbol M$ are positive semidefinite, with
	\begin{align}\label{Eq:MatrixOfMomentsCriterion}
		0\leq& \boldsymbol M=\left(\langle{:} \hat \pi^{m+m'}{:}\rangle\right)_{m,m'=0}^{\lfloor N/2\rfloor}.
	\end{align}
	A violation of this positive semidefiniteness would imply that the measured quantum state is a nonclassical one.

	As an example, let us consider the leading $2\times 2$ minor:
	\begin{align}\label{Eq:2x2QB}
		&\det \begin{pmatrix}
			\langle{:} \hat \pi^{0}{:}\rangle & \langle{:} \hat \pi^{1}{:}\rangle\\
			\langle{:} \hat \pi^{1}{:}\rangle & \langle{:} \hat \pi^{2}{:}\rangle
		\end{pmatrix}
		=\det \begin{pmatrix}
			 1 & \frac{\langle c\rangle}{N} \\
			\frac{\langle c\rangle}{N} & \frac{\langle c^2\rangle-\langle c\rangle}{N(N-1)}
		\end{pmatrix}\\
		=&\frac{N(\langle c^2\rangle-\langle c\rangle^2)-\langle c\rangle(N-\langle c\rangle)}{N^2(N-1)}
		=\frac{\langle c\rangle(N-\langle c\rangle)}{N^2(N-1)}Q_{\rm B}.\nonumber
	\end{align}
	This minor is negative if and only if $Q_{\rm B}<0$.
	Hence, from the general matrix of moment criteria, the second order already represents the binomial $Q_{\rm B}$ parameter.

\begin{figure}[ht]
		\includegraphics[width=8cm]{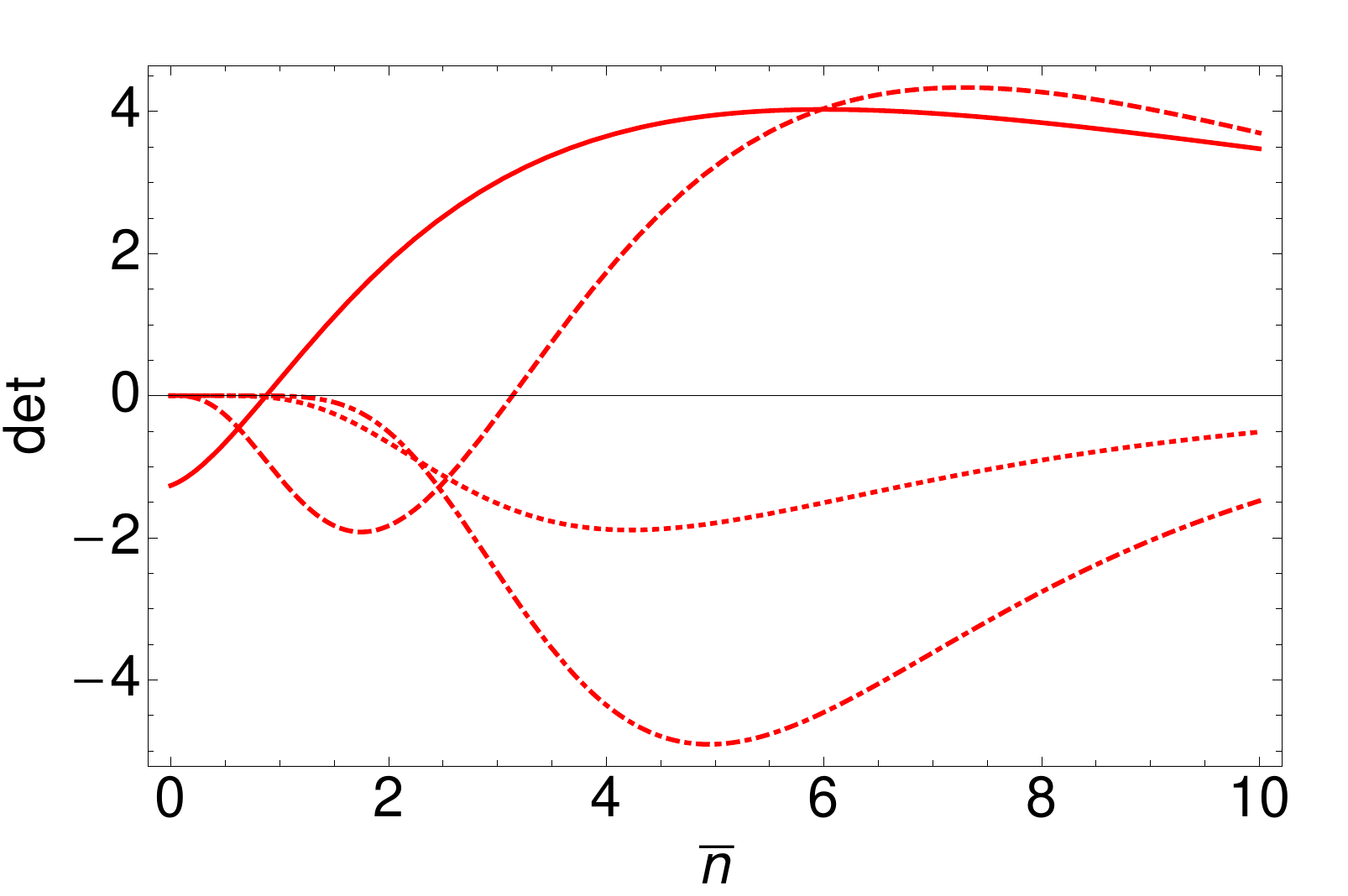}
		\caption{ (Color online)
			A detector with $N=8$ diodes with a quantum efficiency $\eta=0.9$ is considered.
			The single photon added thermal state is parametrized by $\bar n$.
			The curves show:
			the leading principle $2\times2$ minor as a solid line (scaling $10^2$);
			the leading principle $3\times3$ minor as a dashed line (scaling $10^5$);
			the leading principle $4\times4$ minor as a dotted line (scaling $10^8$);
			and the leading principle $5\times5$ minor as a dot-dashed line (scaling $10^{13}$).
		}\label{Fig:HigherOrder}
	\end{figure}

	Higher orders of moments allow a detection beyond this second order criteria.
	In Fig.~\ref{Fig:HigherOrder}, we consider the identification of quantum properties for a single photon added thermal state~\cite{AT92,ZPB07},
	\begin{align}
		\frac{\hat a^\dagger\hat\rho_{\rm th}\hat a}{{\rm tr}\left(\hat a^\dagger\hat\rho_{\rm th}\hat a\right)}=\frac{1}{(\bar n+1)^2}\sum_{n=1}^\infty n\left(\frac{\bar n}{\bar n+1}\right)^{n-1}|n\rangle\langle n|.
	\end{align}
	The parameter $\bar n$ denotes the mean photon number of the initial thermal state $\hat\rho_{\rm th}$.
	Since the $m$th moments scales roughly with $1/N^m$, we scaled the individual minors.
	We see in Fig.~\ref{Fig:HigherOrder}, that different minors violate the positive semi-definiteness condition in Eq.~\eqref{Eq:MatrixOfMomentsCriterion} for different values of $\bar n$.
	Especially for regions where the $Q_{\rm B}$ parameter (leading principal $2\times2$ minor) fails to verify nonclassicality, we demonstrate the advantage of higher order correlations for the identification of quantum properties.

\section{Cross correlations}\label{Sec:3-1}
	Now, we may consider the detection of light with more than one of such detector systems, $D>1$.
	This means, that we have $D$ detectors at different positions; cf. Fig~\ref{Fig:MultiDet}.
	The individual detector $d\in\{1,\ldots, D\}$ consists itself of $N_d$ on-off diodes.
	We readily obtain the joint click counting statistics as
	\begin{align}
		c_{k_1,\ldots,k_D}=& \left\langle{:}\bigotimes_{d=1}^D \frac{N_d!}{k_d!(N_d-k_d)!}\left(\exp\left[-\frac{\eta_d\hat n_d}{N_d}\right]\right)^{N_d-k_d}\right.\nonumber\\
		&\left.\phantom{\bigotimes_{d=1}^D} \times \left(\hat 1_d-\exp\left[-\frac{\eta_d\hat n_d}{N_d}\right]\right)^{k_d} {:}\right\rangle,\label{Eq:JointCCSlarge}
	\end{align}
	with $k_d\in\{0,\ldots,N_d\}$ and the individual quantum efficiencies $\eta_d$.
	For an efficient formulation, it is useful to recall the multi-index notion.
	That is, $\vec k!=k_1!\cdot\ldots\cdot k_D!$, $|\vec k|=k_1+\ldots+k_D$, and $\vec x^{\vec k}=x_1^{k_1}\cdot\ldots\cdot x_D^{k_D}$ (for any multi-index $\vec k\in\mathbb N^D$).
	Moreover multi-indices are ordered as $\vec k< \vec k'$, i.e., $k_1< k'_1,\ldots,k_D<k'_D$.
	For our purpose it is useful to consider a natural generalization.
	If $\hat x_1,\ldots,\hat x_D$ are given operators, we define
	\begin{align*}
		\vec{\hat x}^{\vec k}=\hat x_1^{k_1}\otimes\ldots\otimes\hat x_D^{k_D}.
	\end{align*}
	Having this notion in mind, we get together with
	\begin{align}
		\hat\pi_d=\hat 1_d-\exp\left[-\tfrac{\eta_d\hat n_d}{N_d}\right],
	\end{align}
	a joint click counting statistics in Eq.~\eqref{Eq:JointCCSlarge} in the form,
	\begin{align}\label{Eq:JointCCScompact}
		c_{\vec k}=\left\langle{:}\frac{\vec N!}{\vec k!\left(\vec N-\vec k\right)!}\vec{\hat\pi}^{\vec k}\left(\vec{\hat 1}-\vec{\hat\pi}\right)^{\vec N-\vec k}{:}\right\rangle.
	\end{align}
	Similar to the approach for $D=1$, we get the multidimensional generating function for any $D$ as
	\begin{align}
		g(\vec z)=\left\langle{:}\bigotimes_{d=1}^D\left(z_d\hat\pi_d+[\hat 1_d-\hat\pi_d]\right)^{N_d}{:}\right\rangle.
	\end{align}
	Analogously, the $\vec m$th multi-dimensional factorial moment is described by
	\begin{align}
		\frac{\left(\vec N-\vec m\right)!}{\vec N!}\left.\frac{\partial^{|\vec m|} g(\vec z)}{\partial \vec z^{\,\vec m}}\right|_{\vec z=\vec 1}=\langle {:}\vec{\hat \pi}^{\vec m}{:}\rangle,
	\end{align}
	directly in terms of the measured joint click-counting statistics $c_{\vec k}$.
	A nonclassicality condition is given by the violation of the positive semidefiniteness of the matrix of moments,
	\begin{align}\label{Eq:SystemOfMatrixOfMoments}
		\boldsymbol M=\left(\langle{:}\vec{\hat\pi}^{\vec m+\vec m'}{:}\rangle\right)_{\vec m,\vec m'=\vec 0}^{\lfloor \vec N/2\rfloor}\geq 0.
	\end{align}

	In the following, let us restrict to the scenario $D=2$.
	Hence, we have a joint click-counting distribution of the form,
	\begin{align}
		c_{k_1,k_2}=&\frac{N_1!}{k_1!(N_1-k_1)!}\frac{N_2!}{k_2!(N_2-k_2)!}\\
		\nonumber& \times\left\langle{:}\hat\pi_1^{k_1}(\hat 1_1-\hat\pi_1)^{N_1-k_1} \hat\pi_2^{k_2}(\hat 1_2-\hat\pi_2)^{N_2-k_2}{:}\right\rangle,
	\end{align}
	with $\hat\pi_d=\hat 1_d-\exp\left[-\eta_d\hat n_d/N_d\right]$ for $d=1,2$.
	Let us also note that we used the commonly accepted notion $\hat\pi_1\hat\pi_2$ instead of $\hat\pi_1\otimes\hat\pi_2$ for arbitrary powers of these operators.
	In this case, the matrix of moments reads as
	\begin{widetext}
	\begin{align}
		\boldsymbol M=\begin{pmatrix}
			1 & \langle{:}\hat\pi_1{:}\rangle & \langle{:}\hat\pi_2{:}\rangle & \langle{:}\hat\pi_1^2{:}\rangle & \langle{:}\hat\pi_1\hat\pi_2{:}\rangle & \langle{:}\hat\pi_2^2{:}\rangle & \dots\\
			\langle{:}\hat\pi_1{:}\rangle & \langle{:}\hat\pi_1^2{:}\rangle & \langle{:}\hat\pi_1\hat\pi_2{:}\rangle & \langle{:}\hat\pi_1^3{:}\rangle & \langle{:}\hat\pi_1^2\hat\pi_2{:}\rangle & \langle{:}\hat\pi_1\hat\pi_2^2{:}\rangle & \dots\\
			\langle{:}\hat\pi_2{:}\rangle & \langle{:}\hat\pi_1\hat\pi_2{:}\rangle & \langle{:}\hat\pi_2^2{:}\rangle & \langle{:}\hat\pi_1^2\hat\pi_2{:}\rangle & \langle{:}\hat\pi_1\hat\pi_2^2{:}\rangle & \langle{:}\hat\pi_2^3{:}\rangle & \dots\\
			\vdots & \vdots & \vdots & \vdots & \vdots & \vdots & \ddots
		\end{pmatrix}.
	\end{align}
	\end{widetext}

	Now, we consider the principal second order minor
	\begin{align}\label{Eq:2ndOrdCrossPre}
		&\det\begin{pmatrix}
		      \langle{:} \hat\pi_1^0\hat\pi_2^0 {:}\rangle & \langle{:} \hat\pi_1^1\hat\pi_2^0 {:}\rangle & \langle{:} \hat\pi_1^0\hat\pi_2^1 {:}\rangle \\
		      \langle{:} \hat\pi_1^1\hat\pi_2^0 {:}\rangle & \langle{:} \hat\pi_1^2\hat\pi_2^0 {:}\rangle & \langle{:} \hat\pi_1^1\hat\pi_2^1 {:}\rangle \\
		      \langle{:} \hat\pi_1^0\hat\pi_2^1 {:}\rangle & \langle{:} \hat\pi_1^1\hat\pi_2^1 {:}\rangle & \langle{:} \hat\pi_1^0\hat\pi_2^2 {:}\rangle
		\end{pmatrix}\\
		=&\langle{:} (\Delta\hat\pi_1)^2{:}\rangle\langle{:} (\Delta\hat\pi_2)^2{:}\rangle-\langle{:} (\Delta\hat\pi_1)(\Delta\hat\pi_2){:}\rangle^2,\nonumber
	\end{align}
	with the variances ($d=1,2$) and the moments,
	\begin{align}
		\langle{:}\hat \pi_d{:}\rangle=&\frac{1}{N_d}\sum_{k_1=0}^{N_1}\sum_{k_2=0}^{N_2} k_d c_{k_1,k_2},\\
		\langle{:}\hat \pi_d^2{:}\rangle=&\frac{1}{N_d(N_d-1)}\sum_{k_1=0}^{N_1}\sum_{k_2=0}^{N_2} k_d(k_d-1) c_{k_1,k_2},\\
		\langle{:}\hat \pi_1\hat\pi_2{:}\rangle=&\frac{1}{N_1N_2}\sum_{k_1=0}^{N_1}\sum_{k_2=0}^{N_2} k_1k_2 c_{k_1,k_2}.
	\end{align}
	This criterion relates the covariance matrix of the joint click counting statistics with a nonclassicality criterion for the scenario in Fig.~\ref{Fig:MultiDet}.

	Let us emphasize again, that for all classically correlated light fields holds 
	\begin{align}\label{Eq:2ndOrdCross}
		\langle{:} (\Delta\hat\pi_1)^2{:}\rangle\langle{:} (\Delta\hat\pi_2)^2{:}\rangle\geq\langle{:} (\Delta\hat\pi_1)(\Delta\hat\pi_2){:}\rangle^2.
	\end{align}
	Note the equivalence between minors of $2\times 2$ matrices and the Cauchy-Schwarz inequality:
	\begin{align}
		\det\begin{pmatrix}
			\langle \phi |\phi\rangle & \langle \phi |\phi'\rangle\\
			\langle \phi' |\phi\rangle & \langle \phi' |\phi'\rangle
		\end{pmatrix}=\langle \phi|\phi\rangle\langle\phi'|\phi'\rangle-|\langle \phi |\phi'\rangle|^2\geq0,
	\end{align}
	for some $|\phi{}^{(}{}'{}^{)}\rangle=\hat L{}^{(}{}'{}^{)}|\psi\rangle$.
	Since $|{\rm Im}\langle \phi |\phi'\rangle|\leq |\langle \phi |\phi'\rangle|$, we get for $\hat L{}^{(}{}'{}^{)}=\Delta\hat K{}^{(}{}'{}^{)}=\hat K{}^{(}{}'{}^{)}-\langle\hat K{}^{(}{}'{}^{)}\rangle$, with $\hat K$ and $\hat K'$ being two Hermitian operators, a relation to uncertainty principles,
	\begin{align}
		\langle \phi|\phi\rangle\langle\phi'|\phi'\rangle &\geq |\langle \phi |\phi'\rangle|^2\\
		\nonumber\Rightarrow\, \langle \psi|(\Delta \hat K)^2|\psi\rangle\langle \psi|(\Delta \hat K')^2|\psi\rangle
		&\geq \left|\frac{1}{2{\rm i}}\langle \psi|[\hat K,\hat K']|\psi\rangle\right|^2.
	\end{align}
	Hence, all $2\times 2$ criteria may be formulated as violations of classical Cauchy-Schwarz inequalities or uncertainty relations.
	Thus, a violation of the normally ordered cross-correlation condition in Eq.~\eqref{Eq:2ndOrdCross} identifies correlations which are not present in classical light fields.
	For higher order minors, one gets a manifold of nonclassicality conditions which may be successively tested to identify different orders of quantum correlations.

\begin{figure}[ht]
		\includegraphics[width=8cm]{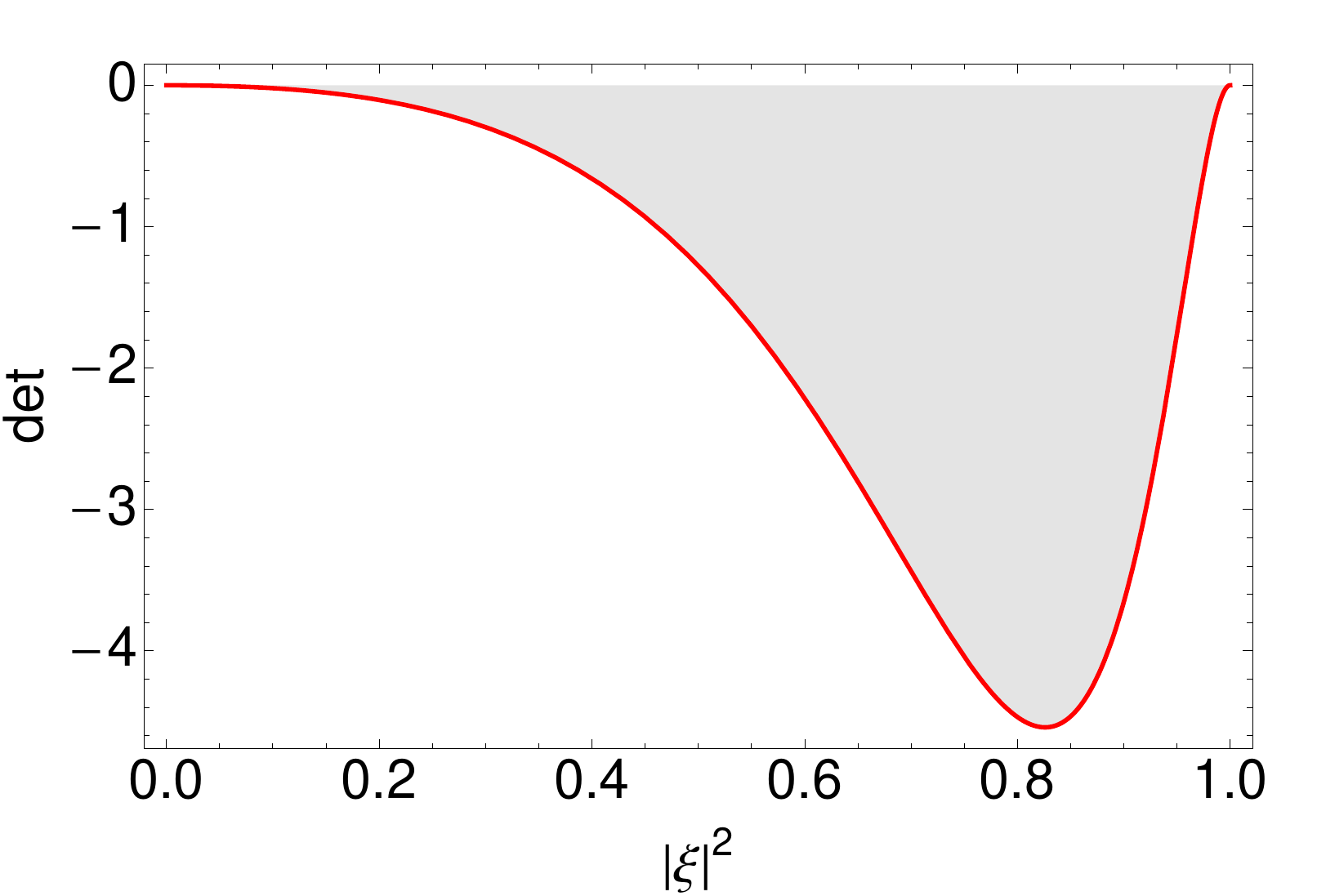}
		\caption{ (Color online)
			The correlation minor in Eq.~\eqref{Eq:2ndOrdCrossPre} (scaled by $10^3$) is plotted for a two-mode squeezed-vacuum state.
			Each of the systems consists of $N_1=N_2=4$ diodes with a quantum efficiency of $\eta_1=\eta_2=0.8$.
		}\label{Fig:TMSV2Det}
	\end{figure}

	In Fig.~\ref{Fig:TMSV2Det}, we consider a correlation measurement according to Eq.~\eqref{Eq:2ndOrdCrossPre} for a two-mode squeezed-vacuum state,
	\begin{align}
		|\xi\rangle=\sqrt{1-|\xi|^2}\sum_{n=0}^\infty\xi^n\, |n\rangle_{1}\otimes|n\rangle_{2},
	\end{align}
	with the squeezing parameter $\xi$ ($|\xi|<1$).
	Differently to the previous example, we measure this state with two click detector systems.
	The part $|n\rangle_{1(2)}$ is measured with the first(second) detector at the position $\boldsymbol r_1(\boldsymbol r_2)$, respectively.
	For all values of squeezing, $0<|\xi|^2<1$, we observe a nonclassical correlation between the subsystems being directly based on the measured joint click counting statistics.
	Let us also point out that the nonclassicality is determined for a non-unit quantum efficiency ($\eta_1,\eta_2<1$) for both detector systems.

\section{Detection of Space Time Dependent Fields}\label{Sec:2}
	In this section, we aim to generalize the description of the click counting statistics to more general scenarios.
	This will include a multi-mode description and a nonlinear response to photon absorption processes.
	We will prove that this will not affect the type of statistics, i.e., the binomial structure.

\subsection{General time-dependent radiation}
	Let us start with a single diode, i.e., $D=1$ and $N=1$.
	The photoelectric detection of light is described by a bulk of atoms.
	Each of these atoms may eject a photoelectron, when it is illuminated with some light.
	These electrons produce -- by an avalanche process -- a current, which is measured.
	For an introduction to photoelectric measurements, we refer to Refs.~\cite{KK64} and~\cite{VW06}~(chapter 6).
	The initial interaction between the atoms and the light field yields a detector response given by $\hat \Gamma(t,\Delta t)$,
	for a measurement within the time interval $[t,t+\Delta t]$.
	This gives a probability for the no-click event for a multimode coherent light field $|\{\alpha_\mu\}\rangle$ of the form,
	\begin{align}
		p(\{\alpha_\mu\};[t,t+\Delta t])=\langle\{\alpha_\mu\}|\NTO \exp\left[ -\hat\Gamma(t,\Delta t) \right]\NTO|\{\alpha_\mu\}\rangle,
	\end{align}
	where $\NTO\ldots\NTO\equiv\mathcal T{:}\ldots{:}$ denotes the normal- and time-ordering prescription and each $\alpha_\mu$ denotes the coherent amplitude of the mode described by the index $\mu$.

	According to the derivation in Ref.~\cite{SVA12a}, it can be readily shown for a quantum state in $P$~representation,
	\begin{align}\label{Eq:GlauberSudarshanP}
		\hat\rho=\int \{{\rm d}^2\alpha_\mu\}\,P(\{\alpha_\mu\})\,|\{\alpha_\mu\}\rangle\langle\{\alpha_\mu\}|,
	\end{align}
	that the time dependent click counting statistics for $N$ on-off detectors is given in the form,
	\begin{align}
		\nonumber c_k(t,\Delta t)=\int& \{{\rm d}^2\alpha_\mu\}\,P(\{\alpha_\mu\})\frac{N!}{k!(N-k)!}\\
		&\times\left(p(\left\{\frac{\alpha_\mu}{\sqrt N}\right\};[t,t+\Delta t])\right)^{N-k}\\
		\nonumber &\times\left(1-p(\left\{\frac{\alpha_\mu}{\sqrt N}\right\};[t,t+\Delta t])\right)^{k},
	\end{align}
	where $k\in\{0,\ldots,N\}$ denotes the number of clicks in the time interval $[t,t+\Delta t]$.
	Note that each of the $N$~diodes is equally illuminated with a fraction of $1/N$ of the initial field intensity. 
	We define the operator $\hat\Gamma^{(N)}(t,\Delta t)$ by the relation
	\begin{align}
		\nonumber &\langle\{\alpha_\mu\}|\NTO\! \exp\!\left[ -\hat\Gamma^{(N)}(t,\Delta t) \right]\!\NTO|\{\alpha_\mu\}\rangle\\
		=&p\left(\left\{\frac{\alpha_\mu}{\sqrt N}\right\};[t,t+\Delta t]\right).
	\end{align}
	In the previously considered, simple cases, we had $\hat \Gamma=\eta\hat n$ and $\hat \Gamma^{(N)}=\eta\hat n/N$.
	In general, we get for $k=0,\ldots,N$,
	\begin{align}
		\nonumber c_k(t,\Delta t)=&\left\langle\NTO \frac{N!}{k!(N-k)!} \left(\exp\!\left[ -\hat\Gamma^{(N)}(t,\Delta t) \right]\!\right)^{N-k}\right.\\
		&\times\left.\left(\hat 1-\exp\!\left[ -\hat\Gamma^{(N)}(t,\Delta t) \right]\!\right)^{k}\NTO\right\rangle.
	\end{align}
	Now, the space time dependent detection, with $D$ on-off detector systems is given by
	\begin{align}
		\nonumber c_{\vec k}(\vec t,\Delta \vec t)=& \left\langle\NTO \frac{\vec N!}{\vec k!(\vec N-\vec d)!}\right.\\
		\label{Eq:TIMEsoso} &\bigotimes_{d=1}^N\left(\exp\left[ -\hat\Gamma^{(N_d)}_d(t_d,\Delta t_d)\right]\right)^{N_d-k_d}\\
		\nonumber &\times\left.\left(\hat 1_d-\exp\left[ -\hat\Gamma^{(N_d)}_d(t_d,\Delta t_d)\right]\right)^{k_d}\NTO\right\rangle.
	\end{align}
	Here, $\hat \Gamma_d(t_d,\Delta t_d)$ denotes the response function of a single on-off detector at a position $\boldsymbol r_d$.

	A generalized, time dependent formulation of the matrix of moments is possible, cf.~\cite{V08,MBWLN10}.
	In this case one replaces the moments in \eqref{Eq:SystemOfMatrixOfMoments} with time dependent moments $\langle\NTO \hat\pi_1^{m_1+m_1'}[t_1]\otimes\ldots\otimes \hat\pi_D^{m_D+m_D'}[t_D]\NTO\rangle$, with
	\begin{align}
		\hat\pi_d[t_d]=\hat 1_d-\exp\left[ -\hat\Gamma^{(N_d)}_d(t_d,\Delta t_d)\right],
	\end{align}
	for $d=1,\ldots, D$.
	For simplicity reasons, we omitted the dependence on $\Delta t_1,\ldots,\Delta t_D$.
	Again, the simplest second order criterion results in
	\begin{align}\label{Eq:2ndOrdCrossNTO}
		\langle\!\NTO\! (\Delta\hat\pi_1[t_1])^2\!\NTO\!\rangle\langle\!\NTO\! (\Delta\hat\pi_2[t_2])^2\!\NTO\!\rangle\geq\langle\!\NTO\! (\Delta\hat\pi_1[t_1])(\Delta\hat\pi_2[t_2])\!\NTO\!\rangle^2.
	\end{align}
	Contrary to the expression in Eq.~\eqref{Eq:2ndOrdCross}, this cross correlation may be used to identify temporal correlations and not only spatial ones.
	Altogether, we get a hierarchy of nonclassicality criteria from all minors of the time dependent matrix of moment for correlated measurements at space time points $(\boldsymbol r_1,t_1)\ldots(\boldsymbol r_D,t_D)$.

\begin{figure}[ht]
		\includegraphics[width=8cm]{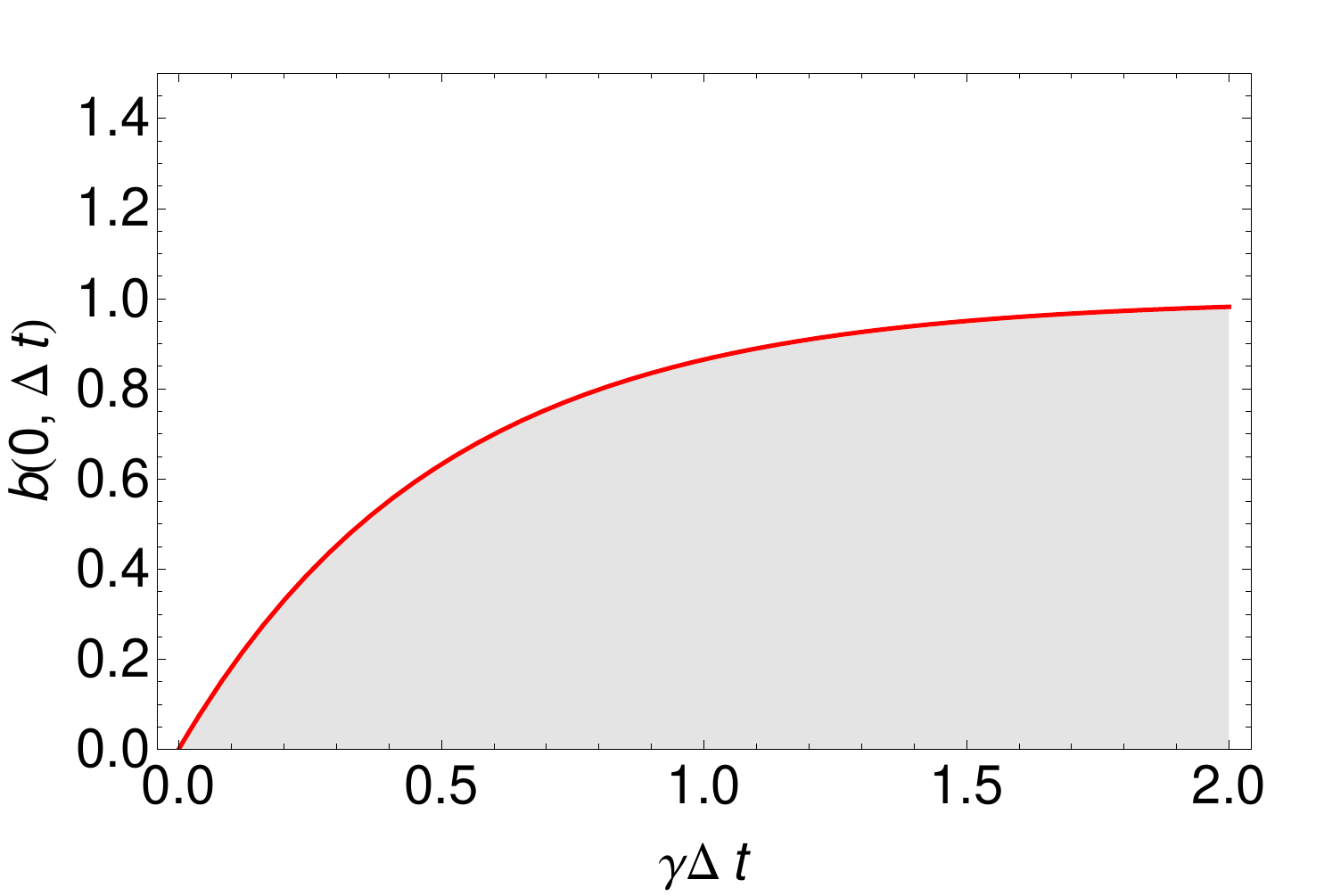}
		\includegraphics[width=8cm]{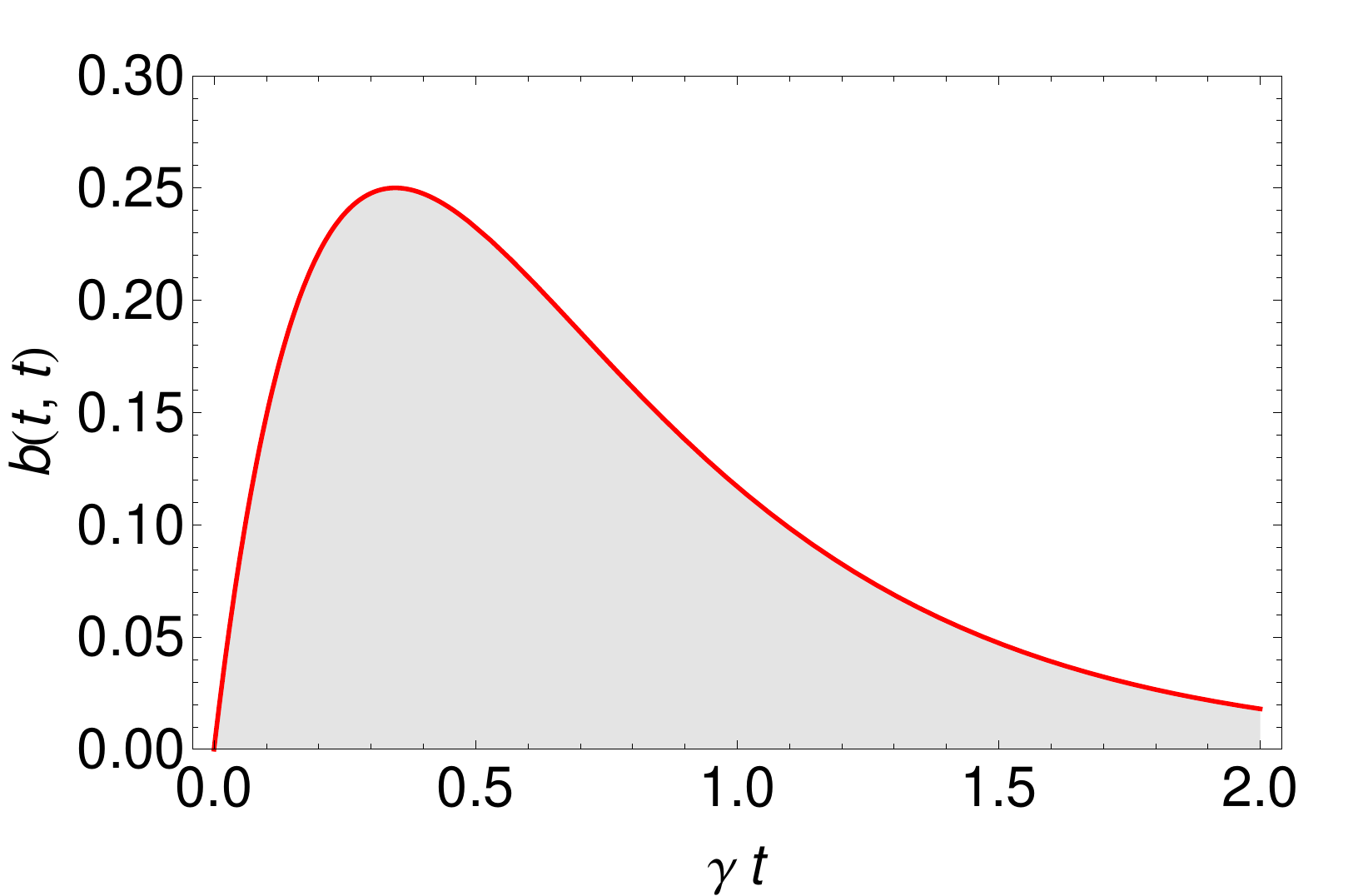}
		\caption{ (Color online)
			We plot the correlation function $b$, measured with a click detector system, depending on the measurement time $t$ and measurement duration $\Delta t$.
			We observe an initially rising behavior of these functions for increasing measurement times, which relates this scenario to the photon-antibunching measurement with standard detectors; cf.~\cite{KDM77,ZM90}.
		}\label{Fig:TimeCorrelations}
	\end{figure}

	As an example of dynamics, we consider the decay of a single photon into a bath.
	For $t=0$ we have a state $|1\rangle_a\otimes|0\rangle_c$ and bosonic operators $\hat a$ for the field and $\hat c$ for the bath.
	The evolution in Heisenberg picture may be ($\gamma>0$),
	\begin{align}
		\hat a(t)=\exp(-\gamma t)\hat a+\sqrt{1-\exp(-2\gamma t)}\hat c.
	\end{align}
	Hence, we have a single photon in the electric field with a probability $\exp(-2\gamma t)$.
	Moreover, a pointlike, broad-band detector (at position $\boldsymbol r$) is described through the function,
	\begin{align}\label{Eq:IntensityCorrelationFunction}
		\hat\Gamma(t,\Delta t)=\xi\int_{t}^{t+\Delta t}{\rm d}\tau\, \hat{\boldsymbol E}^{(-)}(\boldsymbol r,\tau)\cdot\hat{\boldsymbol E}^{(+)}(\boldsymbol r,\tau),
	\end{align}
	where $\xi$ characterizes the dipole transition of the atoms in the detector medium and $\Delta t$ the duration of the measurement~\cite{VW06}.
	The electric field is $\hat{\boldsymbol E}^{(+)}(\boldsymbol r,t)=\boldsymbol E(\boldsymbol r)\hat a(t)$.
	Here, the response in the $N$ on-off detector scenario is given by $\hat \Gamma^{(N)}=\hat \Gamma/N$.
	In this scenario of a single photon state, we get a second order matrix of moment in the form ($D=1$),
	\begin{align}\label{Eq:ANTIbunchING}
		\det\begin{pmatrix}
			\langle\NTO\hat 1\NTO\rangle & \langle\NTO\hat \pi[t]\NTO\rangle \\
			\langle\NTO\hat \pi[t]\NTO\rangle & \langle\NTO\hat \pi^2[t]\NTO\rangle
		\end{pmatrix}
		=&-\frac{2\langle\NTO\hat\Gamma(t)\NTO\rangle}{N^2(N-1)}\\
		=&-\frac{\xi|\boldsymbol E(\boldsymbol r)|^2}{2\gamma N^2(N-1)}b(t,\Delta t),\nonumber
	\end{align}
	which is described by the properly normalized function,
	\begin{align}
		b(t,\Delta t)=\exp[-2\gamma t]\left(1-\exp[-2\gamma \Delta t]\right).
	\end{align}
	In Fig.~\ref{Fig:TimeCorrelations}, we plot $b(t,\Delta t)$.
	Whenever this function is positive, we have a negative minor in Eq.~\eqref{Eq:ANTIbunchING}.
	With an increasing measurement time, $\Delta t\to\infty$ , we approach the value 1.
	Additionally, we have $\lim_{t\to\infty}b(t,\Delta t)=0$.
	Hence the correlations would disappear together with the decay of the photon.
	Note that a coherent light field is described by a constant function.
	
\subsection{Multimode expansion}
	We considered monochromatic light fields in Secs.~\ref{Sec:3} and~\ref{Sec:3-1}, and Refs.~\cite{SVA12a,SVA12b}.
	Let us now consider the relation to multimode systems.
	Analogously to the example in Eq.~\eqref{Eq:IntensityCorrelationFunction}, we may consider a linear, phase-independent detection model.
	This is given by
	\begin{align}
		\hat \Gamma=\sum_\mu \eta_\mu\hat n_\mu,
	\end{align}
	where $\eta_\mu$ is the detection efficiency and $\hat n_\mu=\hat a^\dagger_\mu\hat a_\mu$ is the photon number operator of the mode $\mu$.
	For simplicity, we consider no time dependence -- a generalization is straight forward.
	On the one hand, a multimode coherent state $|\{\alpha_\mu\}\rangle$ yields a click counting statistics,
	\begin{align}
		c_k=& \frac{N!}{k!(N-k)!}\left(\exp\left[ -\frac{\sum_\mu\eta_\mu|\alpha_\mu|^2}{N}\right]\right)^{N-k} \nonumber\\
		&\phantom{\bigotimes_{d=1}^D} \times \left(1-\exp\left[ -\frac{\sum_\mu\eta_\mu|\alpha_\mu|^2}{N}\right]\right)^{k}.
	\end{align}
	On the other hand, a monochromatic state $|\beta\rangle$ gives 
	\begin{align}
		c_k=&  \frac{N!}{k!(N-k)!}\!\left(\!\exp\left[ -\frac{\eta|\beta|^2}{N}\right]\!\right)^{\!N-k}\!
		\!\left(\!1{-}\exp\left[ -\frac{\eta|\beta|^2}{N}\right]\!\right)^{\!k}\!,
	\end{align}
	for $\hat \Gamma=\eta\hat n$.
	If we now consider $\eta|\beta|^2=\sum_\mu\eta_\mu|\alpha_\mu|^2$ for a choice of $\eta$ and $\beta$, we cannot observe a difference between the monochromatic case and the polychromatic one.
	More generally, a multimode state $\hat\rho$ given by the Glauber~Sudarshan $P$~function in Eq.~\eqref{Eq:GlauberSudarshanP}
	cannot be distinguished from a single mode state $\hat\rho'=\int{\rm d}^2\beta\,P'(\beta)|\beta\rangle\langle\beta|$ with
	\begin{align}
		P'(\beta)=\int\{{\rm d}^2\alpha_\mu\}&\,P(\{\alpha_\mu\})\,\delta\left(\eta|\beta|^2-\sum_{\mu}\eta_\mu|\alpha_\mu|^2\right).
	\end{align}
	Hence, without a loss of generality, a monochromatic formulation of the system is always possible, when the assumption of a phase-independent detection process is justified.

\subsection{Nonlinear detector response}
	Let us now consider a nonlinear detector response, i.e., $\hat \Gamma$ depends nonlinearly on $\hat n$; cf., e.g.,~\cite{JA69,FW91}.
	This may happen when a nonlinear process dominates the interaction between the light field and the detector atoms.
	In this case of a single mode field, the no-click probability can be described by
	\begin{align}
		\left\langle{:} \exp\left[ -\hat \Gamma^{(N)} \right]{:}\right\rangle=\left\langle{:} \exp\left[ -f(\hat n/N) \right]{:}\right\rangle,
	\end{align}
	for a nonlinear function $\hat\Gamma=f(\hat n)$, which is assumed to be phase independent.
	Hence, we get for the function $f$ the nonlinear click-counting statistics as
	\begin{align}
		\nonumber c_k(f)=&\left\langle {:}\frac{N!}{k!(N-k)!}\left(\exp\left[-f(\hat n/N)\right]\right)^{N-k} \right.\\
		&\left.\phantom{\times}\times\left(\hat 1-\exp\left[-f(\hat n/N)\right]\right)^{k}{:}\right\rangle.
	\end{align}

	As an example for such a scenario, let us consider a single detector, which is based on $n_0$-photon absorption.
	In this case there is no click, if the number of incident photons is below $n_0$.
	The probability is
	\begin{align}
		\nonumber \left\langle{:} \exp\left[ -\hat \Gamma \right]{:}\right\rangle=&\big\langle|0\rangle\langle 0|+\ldots+|n_0-1\rangle\langle n_0-1|\big\rangle\\
		=&\left\langle{:}\left(\sum_{j=0}^{n_0-1}\frac{\hat n^j}{j!}\right)\exp\left[-\hat n\right]{:}\right\rangle\\
		\nonumber =&\left\langle{:}\exp\left[-\left(\hat n-\log\left[\sum_{j=0}^{n_0-1}\frac{\hat n^j}{j!}\right]\right)\right]{:}\right\rangle.
	\end{align}
	Thus, we have $f(x)=x-\log\left[\sum_{j=0}^{n_0-1}x^j/j!\right]$.

	Let us consider another scenario where the electric current of a photo diode nonlinearly depends on the photon number.
	In general, one can perform a Taylor expansion, $\hat \Gamma=c_0\hat 1+c_1\hat n+c_2\hat n^2+\ldots$, of the phase-independence detector response.
	Hence, the nonlinear intensity dependence for a single term is $f(x)=x^{n_0}$.
	This yields a click-counting statistics in the form,
	\begin{align}
		\nonumber c_k(f)=&\left\langle {:}\frac{N!}{k!(N-k)!}\left(\exp\left[-\left(\tfrac{\hat n}{N}\right)^{n_0}\right]\right)^{N-k}\right.\\
		&\left.\phantom{\times}\times\left(\hat 1-\exp\left[-\left(\tfrac{\hat n}{N}\right)^{n_0}\right]\right)^{k}{:}\right\rangle,
	\end{align}
	for $k=0,\ldots,N$.

\begin{figure}[ht]
		\includegraphics[width=4cm]{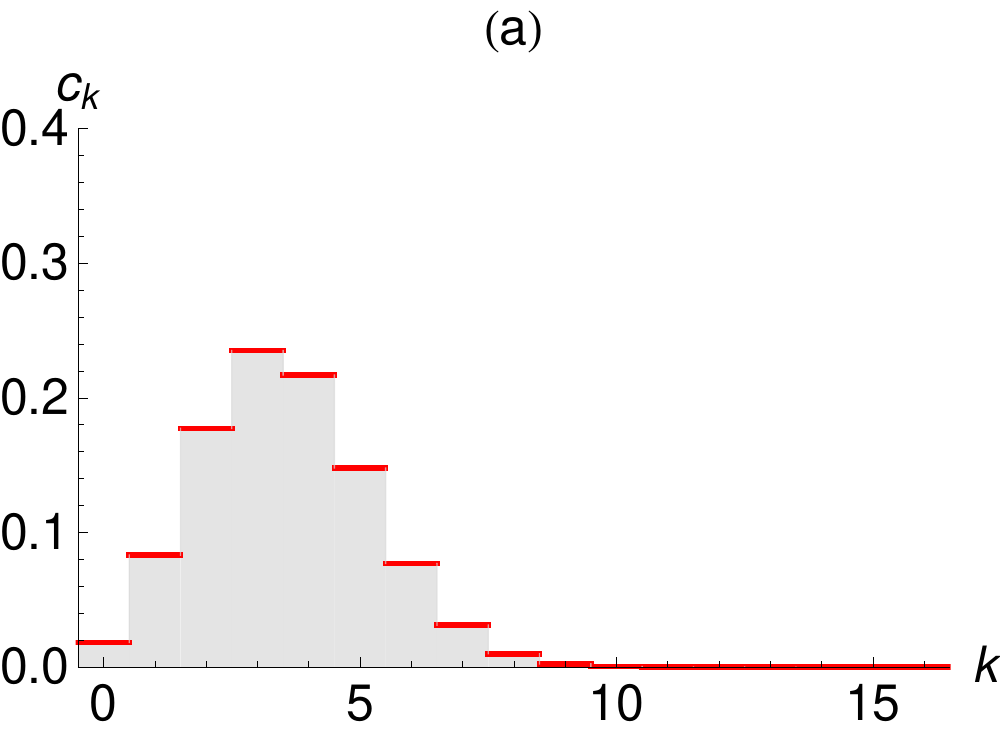}
		\includegraphics[width=4cm]{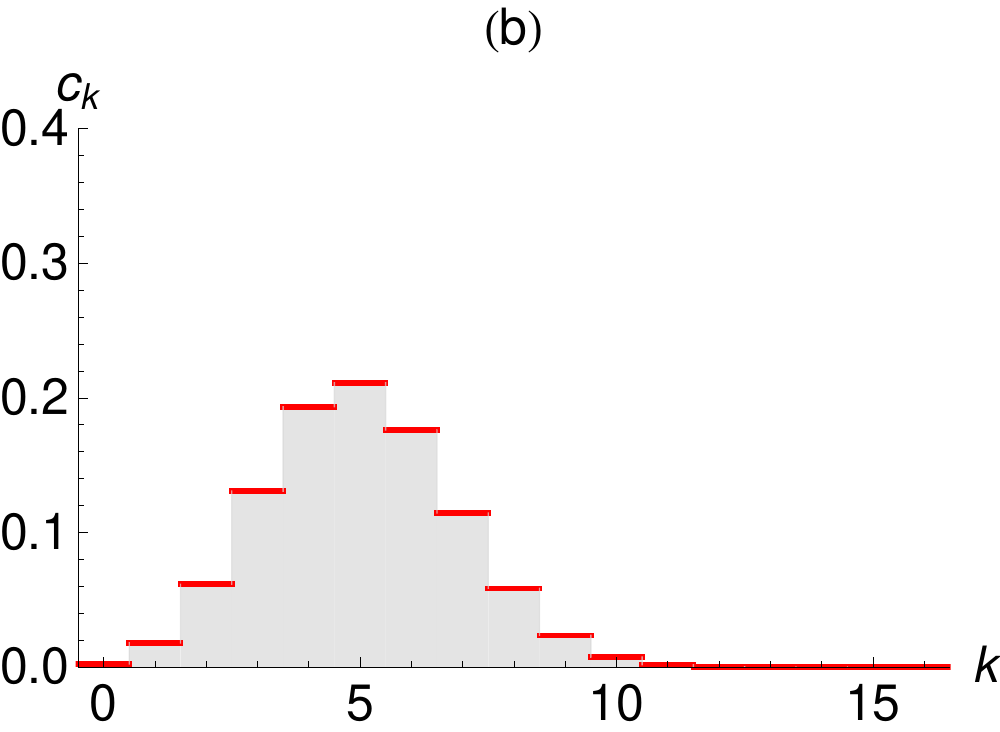}
		\includegraphics[width=4cm]{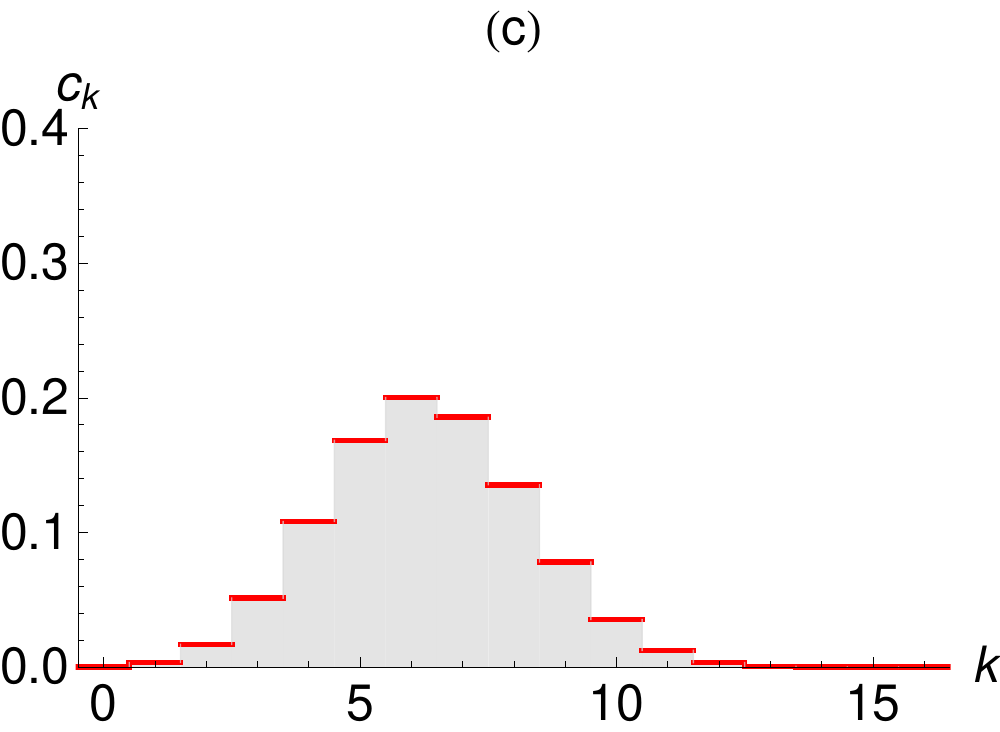}
		\includegraphics[width=4cm]{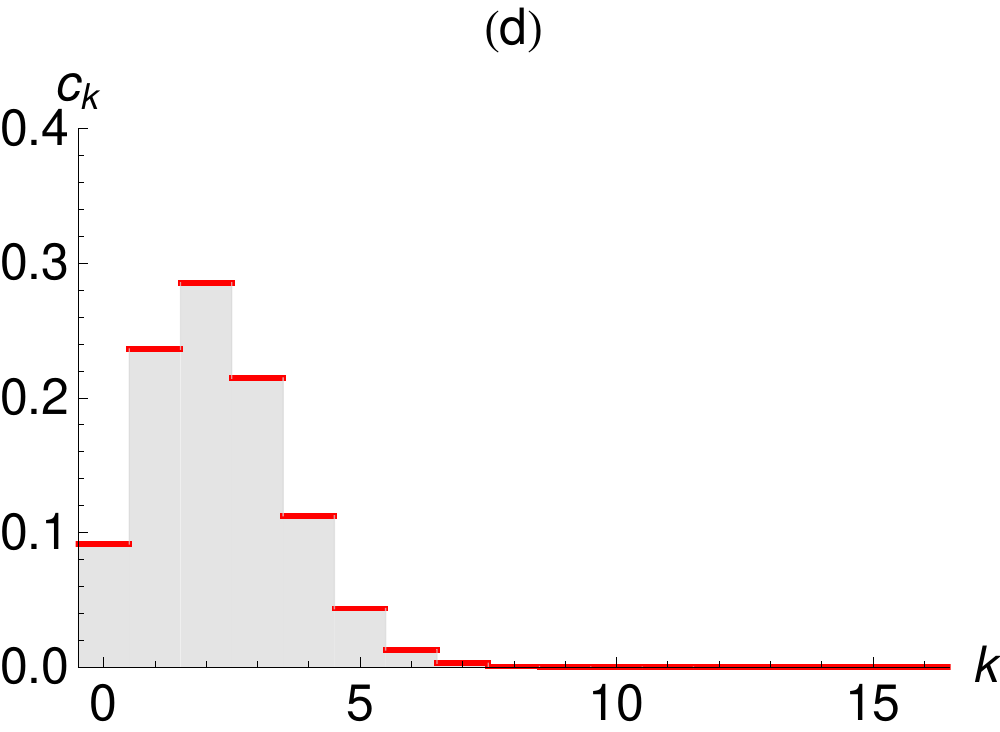}
		\caption{ (Color online)
			The click-counting statistics ($N=16$) is shown for coherent state $\alpha=2$ for different response functions:
			(a) a linear function $f(x)=x$; (b) an affine function $f(x)=x+2$; (c) a quadratic response $f(x)=x+\frac{1}{4}x^2$; and (d) two-photon absorption $f(x)=x-\log[1+x]$.
		}\label{Fig:NonlinResponse}
	\end{figure}

	In Fig.~\ref{Fig:NonlinResponse}, we compare the click statistics of coherent states for different nonlinear functions.
	We observe that the binomial structure of this classical light field is,
	\begin{align}
		c_k(f)=\frac{N!}{k!(N-k)!}(1-p)^{N-k}p^k,
	\end{align}
	with $p=1-\exp[-f(|\alpha|^2/N)]$.
	Compared with a linear response, the nonlinear functions give different probabilities $p$.
	However, the binomial behavior of coherent light is preserved.
	It is also worth mentioning that a function $f_\nu(x)=f(x)+\nu$ yields the description of a dark count rate $\nu$ of the detector; cf.~\cite{SVA12a}.

\begin{figure}[ht]
		\includegraphics[width=8cm]{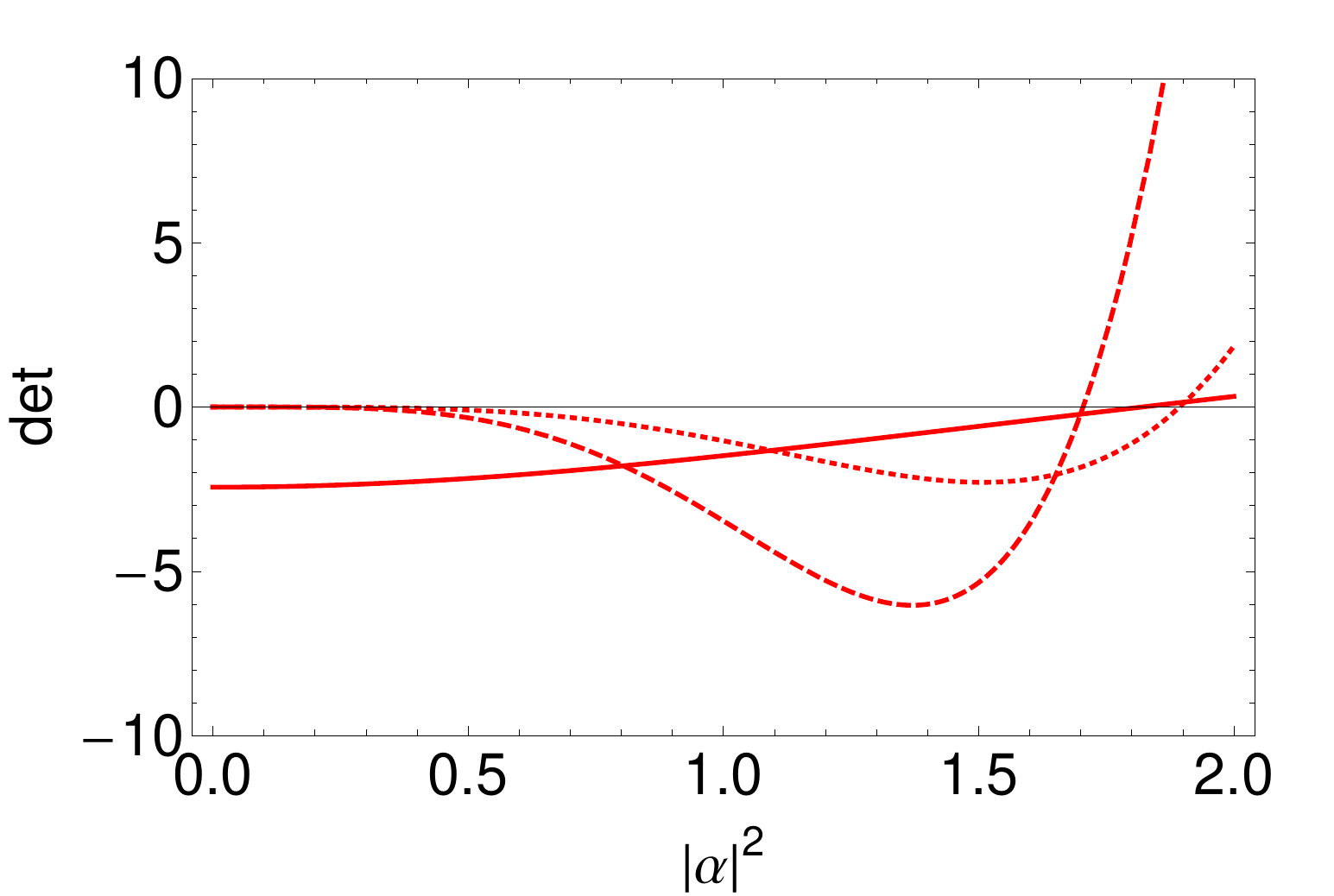}
		\caption{ (Color online)
			The principal leading $2\times 2$ minor ($N=8$) is shown for an odd coherent state depending on $|\alpha|^2$.
			The response functions are $f(x)=x$ (solid, scaling $10^4$), $f(x)=x^3$ (dashed, scaling $10^8$), and $f(x)=x-\log[1+x+\tfrac{1}{2}x^2]$ (dotted, scaling $10^9$).
		}\label{Fig:NonlinResponseEvenOdd}
	\end{figure}

	As an example for a nonclassical state, we consider the odd coherent state,
	\begin{align}
		|\alpha\rangle_{-}=\mathcal N_{-}\left(|\alpha\rangle-|-\alpha\rangle\right),
	\end{align}
	with $\alpha\in\mathbb C\setminus\{0\}$ and the normalization constant $\mathcal N_{-}=[2(1-\exp[-2|\alpha|^2])]^{-1/2}$.
	Here, it is important to recall that the odd coherent state solely contains odd photon numbers in the Fock state expansion.
	Due to the obvious relation $\langle \alpha|{:}\hat n^p{:}|-\alpha\rangle=(-|\alpha|^2)^p\exp[-2|\alpha|^2]$, we get nonlinear moments,
	\begin{align}
		\langle{:}\hat\pi^m{:}\rangle=&\frac{(N-m)!}{N!}\langle{:}\left(\hat 1-\exp\left[-f(\hat n/N)\right]\right)^m{:}\rangle\\
		\nonumber =&\frac{(N-m)!}{N!}\frac{1}{1-\exp[-2|\alpha|^2]}\\
		\nonumber &\times \left[\left(1-\exp\left[-f(|\alpha|^2/N)\right]\right)^m\right.\\
		\nonumber &\phantom{\times\times}\left. -\left(1-\exp\left[-f(-|\alpha|^2/N)\right]\right)^m\exp[-2|\alpha|^2]\right].
	\end{align}
	In Fig.~\ref{Fig:NonlinResponseEvenOdd}, we show the minor $\langle{:}\hat\pi^2{:}\rangle-\langle{:}\hat\pi{:}\rangle^2$ of the matrix of moments, which corresponds to a nonlinear $Q_{\rm B}$ parameter; cf. Eq.~\eqref{Eq:2x2QB}.
	The negativities of the minor identify nonclassical light fields, even if details of the detector response, such as its nonlinearities, are unknown.

\section{Summary and Conclusions}\label{Sec:4}
	In summary, we formulated a hierarchy of conditions for nonclassical correlations measured with on-off detector systems.
	These criteria have been formulated in terms of matrices of moments.
	Each principal minor of this matrix can identify nonclassical quantum states, if it becomes negative.
	We demonstrated that higher order moments can detect quantum effects beyond the second-order criterion.
	Let us stress that these moments are based solely on the measured click-counting statistics;
	a pseudoinversion to the photon statistics is superfluous.
	We generalized these conditions to correlation measurements of joint click counting statistics for multiple detector systems.
	Based on this method, quantum correlations between radiation fields can be revealed.

	We have shown that a generalized detection theory can be applied to systems of on-off detectors.
	The binomial behavior of joint click counting of coherent light is unchanged for different types of response functions, even for a nonlinear detection model.
	The corresponding nonclassicality criteria hold true and can be directly applied.
	Quantum effects can be identified by our method, although some features of the detector -- such as the response behavior -- might be unknown.

	In addition, we outlined that the matrix of moments allows a general space-time-dependent identification of quantum effects.
	For this reason, we studied the description of the click counting statistics for time-dependent measurement processes.
	Altogether, on the basis of our method a more efficient application of multiplexing and array detectors becomes feasible.

\section*{Acknowledgments}
	This work was supported by the Deutsche Forschungsgemeinschaft through SFB 652.
	J.S. gratefully acknowledges financial support from the Oklahoma State University.

\end{document}